\documentstyle[12pt,epsfig]{article}
\newcommand{\be}{\begin{equation}}
\newcommand{\ee}{\end{equation}}

 1
 1
 1

\def\reff{\hang\noindent}

\begin{document}
\vspace*{1.8cm}
  \centerline{\bf COSMIC RAYS AND THE STRUCTURE OF SPACE-TIME}
\vspace{1cm}

\centerline{AURELIO F. GRILLO\footnote{Talk presented by A.F. Grillo.}$^{,a}$ 
and ROBERTO ALOISIO$^{a,b}$}

\vskip 1 truecm
\centerline {\it $^a$ Istituto Nazionale di Fisica Nucleare, 
Laboratori Nazionali del Gran Sasso,}
\centerline {\it  67010 Assergi (L'Aquila), (Italy). }
\centerline {\it $^b$ Dipartimento di Fisica dell'Universit\`a 
di L'Aquila,  67100 L'Aquila, (Italy).}
\vspace{3cm}

\begin{abstract}
Even a fundamental symmetry like Lorentz Invariance is an experimental
fact and must be experimentally verified. We show 
that the study of the interactions of Cosmic Rays  with 
universal diffuse background radiation can provide very 
stringent tests  of this symmetry.
The interactions we consider are the ones characterized by 
well defined energy thresholds whose energy position can be predicted 
on the basis of special relativity. We argue that the experimental 
verification of these thresholds can  
address the physics of supra-Planckian scales. 
\end{abstract}
\vspace{2.0cm}

\section{ Introduction }

Symmetry principles have generally origin from experimental evidence and
therefore their validity has to be verified  to an ever 
increasing degree of precision, or falsified. 
Lorentz Invariance (LI) is no exception,
being based on experimental facts (e.g. the impossibility of verifying
the motion of the Earth from laboratory experiments) and having innumerable
experimentally testable consequences (e.g. constancy of speed of light,
equivalence of physics in different reference frames ....).

In a recent paper (Aloisio et al. 2000) we discussed the possibility 
of using Cosmic Ray (CR) experiments to put very stringent constraints 
on the validity of LI. That this is the case 
can be intuitively motivated in the
following way. Consider  the process giving rise to the Greisen, 
Kuzmin, Zatsepin cut-off (Greisen 1966, Zatsepin and Kuzmin 1966),
{\it i.e.} pion photoproduction. In the terrestrial laboratory the
reaction is $\gamma p \to \pi N $ and has a threshold at $E_{th} 
\approx 100$ MeV for a (proton) target at rest. In the same frame 
the reaction initiated by UHE CRs on Cosmic Microwave Background (CMB)
photons as a target ($p \gamma_{CMB} \to \pi N$) has a threshold of
$\approx 5 \cdot 10^{19}$ eV. In this frame the two reactions appear
very different: in particular a $10^{20}$ eV proton needs only an
extremely tiny fraction of its energy ($ \approx 10^{-23}$) to make a
transition in the final state containing a pion. Obviously there is
nothing misterious in this, LI just implying that the two reactions
are exactly the same taking place in two reference frames in motion with a 
relative Lorentz factor $\gamma \approx 10^{11}$. However it is clear that 
even very small deviations from strict relativistic invariance
are likely to profoundly modify the value of the thresholds and of the
associated absorption cut-offs, in a way in principle experimentally 
verifiable. A similar process, giving rise to an absorption threshold 
for VHE-UHE Cosmic $\gamma$ rays, is $\gamma \gamma_{BCKG} \to e^+ e^-$ where
again $\gamma_{BCKG}$ is a low energy background photon: IR (Stecker 1999), 
microwave (Nikishov 1962, Goldreich and Morrison 1964, 
Gould and Schreder 1966) or radio (Protheroe and Biermann, 1996, 1997). 

On the other hand, it has been been conjectured since the 50's 
(Wheeler 1957) that
strict Lorentz invariance is likely to be profoundly modidified at the 
scales at which quantum grvitational 
effects begin to be relevant: at Planck distances
(times) the topology of space-time may become highly non trivial, 
making the definition of distance essentially imposible thus profoundly 
modifing our picture of the physical world. 

In this talk I present arguments showing, on very general grounds and 
without referring to specific models, that in the processes 
which give rise to   absorption thresholds for  
the propagation of cosmic rays in the Universe (due the onset of 
particle production on universal background radiation)
it is possible to test the 
validity of LI to a very high degree of precision; 
in particular, if deviations from 
LI are ascribed to QG effects, they can be studied down to length scales
orders of magnitude smaller than Planck length,  even using beam 
particles with energy {\it far less} than the Planck mass!

I want however to remark that we {\it do not} advocate violation of LI 
to explain the features of the spectrum of UHE Cosmic Rays (e.g. the 
apparent absence of the GZK cut-off). Our knowledge of the sources of
the highest energy Cosmic and $\gamma$ rays is rather poor, so that we prefer 
to stress the capability of CR experiments to test such a fundamental 
symmetry. However the situation is going to change drastically 
with the new generation of extremely large collecting area
Cosmic Ray Experiments (Cronin 1992), 
or is already changing, for instance due
to the high statistics studies of the VHE spectrum of extragalactic 
$\gamma$-sources like Mkn421 and 501 (see for instance Aharonian et al. 1999,
Guy et al. 2000).

\section{ Absorption thresholds in non-LI world }

As intuitively motivated above, the reactions that can lead to a verification
of LI are those in which VHE and UHE Cosmic and $\gamma$ rays interact
with a universal diffuse (photon) background, that can be the very well known
Cosmic Microwave Background Radiation (energy of maximum radiance
$\epsilon \approx 10^{-3}$ eV), the less known Far Infrared Radiation
($\epsilon \approx 10^{-2}$ eV) or the hypotetical Radio Background
($\epsilon \approx 10^{-7}$ eV). On this diffused radiation 
UHE protons and VHE $\gamma$'s can interact and produce secondary 
particles, mostly $\pi$ in the case of primary protons, $e^+ e^-$ in the case 
of $\gamma$'s. 

In all the cases the processes can be seen as Lorentz boosted from 
terrestrial laboratory with boost factors ranging from $10^7$
to $10^{14}$.
The cross sections for these processes are large
($\approx 10^{-25}$ $cm^2$) so that the energetic particles are 
rapidly degraded in energy with an absorption length of the order of 
tens Kpc for $\gamma$ absorption on CMBR to tens of Mpc for 
protons absorption in CMBR and $\gamma$ absorption on Radio background.
Both  these facts are important quality factors for using
CR' s to test LI. The first gives the range of parameters in which SR 
can be tested, while the second suggests that possible
modifications of these reactions are expected
to produce a (in principle easily) detectable signal. 

To make the test quantitative one needs a parametrization of possible
Lorentz violations
\footnote{for earlier discussions of LI breaking see: 
Kirzhnits and Chechin 1971, 
Gonzalez-Mestres 1997, Coleman and Glashow 1997, Amelino-Camelia et al. 1997.
A very similar approach is in Amelino-Camelia and Piran 2000.}. 
The constancy of speed of light ($c(=1)$ in 
the following) implies the existence of an invariant interval
$ds^2 = dt^2-dx^2$ (=0 for light signals), and that the
norm of any four-vector is invariant: in particular for the 
Energy-Momentum four-vector of a particle of mass $m$ one has the dispersion
relation: 
\be
P_{\mu}P^{\mu}=m^2 = E^2-|\vec p^2|    
\label{eq:l1}
\ee
This is the relation that we modify in order to parametrize violations
of LI.
We follow a  phenomenological approach and we do not refer to any 
specific model (see however next section).
Our guiding  principles are: 

1)
Violations are universal, $i.e.$ do not depend on
particle type;

2)
Preserve rotational invariance;

3) 
Violations are an high energy phenomenon, vanishing at  low 
momenta.
\\
\noindent
Finally we only consider the 
$p << M_P$ range (relevant for the experiments we cosider).
We therefore modify the dispersion relation as
($p=|\vec p|$): 

\be
E^2-p^2=m^2+p^2 f({p\over M}) + 
.....  
\label{eq:l2}
\ee
The dots stand for terms that are subleading in the regime we are considering; 
also in this regime the difference between $E$ and $p$ in the RHS of eq. 2
is a higher order correction. $M$ is a mass parametrizing the violation of LI,
which is expected to be of the order of the Planck mass $M_P$ if violations
originate from quantum gravity effects,
and $f(0)=0$. Under these conditions we can expand $f(p/M)$ and, reabsorbing
numerical coefficients in M, we can write, for the leading correction:

\be
 E^2-p^2 \approx m^2 \pm  p^2({p \over M})
\quad \quad \quad \quad I_{\pm} 
\label{eq:l3}
\ee

\be
E^2-p^2 \approx m^2 \pm  p^2 ({p^2 \over M^2}) 
\quad \quad \quad \quad II_{\pm} 
\label{eq:l4}
\ee
being of first (second) order in the (small) parameter $p/M$.
 
It is worth noticing that we could have modified the Lorentz transformations
of four-momenta

\be
E'=\gamma M F({E+\beta p \over M}); \quad
p'=\gamma M G({\beta E + p \over M}) \quad \quad (F(0)=G(0)=0) 
\label{eq:l5}
\ee
which, with the appropriate choice of F and G con lead to (3) or (4).
In this procedure there is however a large arbitrariety (eq. (5) is not
even the most general) and we do not pursue it here
\footnote{Notice that if $F(\infty),G(\infty) \ne \infty$ eq. (5) 
implies that there is a maximum value for energy (momentum) (Garay 1999).}.

We also  $assume$
Energy-Momentum conservation. This is not granted, if
the violation of LI is associated to a violation of
traslational invariance, and has to be checked
in specific models. However this assumption is legitimate if we want to 
put bounds on the violations of LI from the $observation$ of the
absorption thresholds.

We have now all the ingredients to compute the value of the threshold 
momenta in this new framework. It is important to notice that the 
computation must be performed in a specific frame, which we
take the one in wich the diffuse radiation is isotropic (neglecting the 
motion of the Earth), 
taking into account 
energy-momentum conservation and the modifications of the dispersion 
relations. This leads to algebraic equations (Aloisio et al. 2000)

\begin{figure}[thb]
 \begin{center}
  \mbox{\epsfig{file=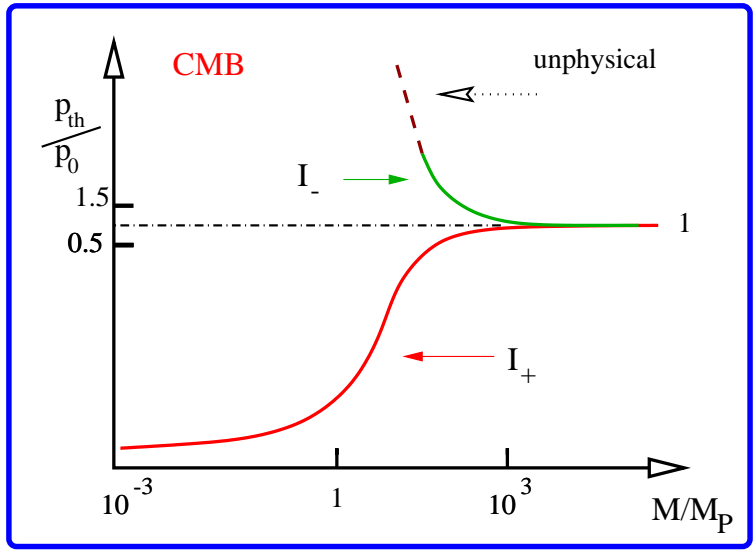,width=10cm}}
  \caption{\em {Qualitative behaviour of the solutions, here for
I-type modifications in the case of $e^+ e^-$ production on CMBR.}}
\end{center}
\end{figure}

\be
\pm \alpha_I x^3+x-1=0
\label{eq:l6}
\ee
\be 
\pm \alpha_{II} x^4+x-1=0
\label{eq:l7} 
\ee
where $x=p_{th}/p_0$ and $p_0$
are the thresholds as $M \to \infty$
\footnote{$ p_0 \approx (m_{\pi}m_p)/(2 \omega)$
for pion production by protons,
$p_0 \approx m_e^2/(2 \omega)$
for pair production by $\gamma$. $\omega$ is the energy of
background photons.},
and (in parentesys the values for pion production)
\begin{eqnarray}
 \alpha_I &=& \frac{p_0^3}{8m_e^2 M}  \quad \quad (
\frac{p_0^3}{m_p^2M}) \\
 \alpha_{II} &=& \frac{3p_0^4}{16m_e^2 M^2}  \quad \quad (
\frac{3p_0^4}{m_p^2 M^2}) \nonumber
\end{eqnarray}

Notice that  if we require that $p_{th} \approx p_0$ then we must have
$\alpha_I$ ($\alpha_{II}$) $<< 1$, which, taking as example the pion production
by protons, implies $M > 10^{14} M_P$ 
($M > 10^{3} M_P$) respectively. Namely, a verification of the LI threshold
momentum value would put a limit on the violation parameter (much) larger
than the Planck mass. 

The qualitative behaviour of the solution is presented in Fig. 1. In case of 
positive modification the threshold moves towards lower values as $M$ moves
away from $\infty$, while it increases, and becomes rapidly unphysical
($i.e.$ negative or complex) for negative modifications; in this case
the reaction becomes kinematically
forbidden.  

In a more quantitative way, in the following table we present the 
values of $p_{th}/p_0$ for pair production, assuming $M=M_P$:
  \begin{center}
\begin{tabular}{|c|c|c|c|}
\hline
\ ~ & Infrared & Microwave & Radio \\
\hline
\ $I_+$ & $\approx 1$ & $0.06$ & $5\cdot 10^{-7}$ \\
\hline
\ $I_-$ & \it no solution  & {\it no solution} & {\it no solution} \\
\hline
\ $II_+$ & $\approx 1$ & $1$ & $2\cdot 10^{-3}$ \\
\hline
\ $II_-$ & $\approx 1$ & $\approx 1$ & {\it no solution} \\
\hline
\end{tabular}
\end{center} 
Notice however that, for positive modifications the
process $\gamma \to e^+ e^-$ becomes allowed, giving rise to absorption
even in absence of target (Coleman and Glashow 1997).

If we leave M as free parameter, {\it assuming} experimental 
verification of thresholds (with a $100 \%$ uncertainty in
momentum determination) we have
\begin{center}
\begin{tabular}{|c|c|c|c|}
\hline
\ ~ & Infrared & Microwave & Radio \\
\hline
\ $I_{+}$ & ${M \over M_P} \geq 0.2$ & ${M \over M_P} 
\geq 800 $ & ${M \over M_P}\geq 10^{18}$  \\
\hline
\ $I_-$ & ${M \over M_P} \geq 6  $ & ${M \over M_P} \geq 3 \cdot 10^4  $ 
& ${M \over M_P} \geq 
8 \cdot 10^{19}  $ \\
\hline
\ $II_{+}$ & $({M \over M_P}\geq 3\cdot 10^{-8} )$ & 
$({M \over M_P} \geq 7 \cdot 10^{-6} )$ & ${M \over M_P}\geq 10^5$ \\
\hline
\ $II_{-}$ & $({M \over M_P}\geq 3\cdot 10^{-7} )$ & 
$({M \over M_P} \geq  10^{-4} M_P)$ & $M\geq 10^6$ \\
\hline
\end{tabular}
\end{center} 

Analogously for pion production (the process giving rise to the GZK
cut-off), for $M=M_P$
\begin{center}
\begin{tabular}{|c|c|}
\hline
\ ~ & GZK \\
\hline
\ $I_+$ & $2\cdot 10^{-5}$ \\
\hline
\ $I_-$ & {\it no solution} \\
\hline
\ $II_+$ & $0.02$ \\
\hline
\ $II_-$ & {\it no solution} \\
\hline
\end{tabular}
\end{center}
Also in this case for positive modifications the process $p \to \pi N$
becomes kinematically allowed.
If we leave M as free parameter, assuming experimental 
verification of thresholds (again within a factor 2 in energy) \\
\begin{center}
\begin{tabular}{|c|c|}
\hline
\ ~ & GZK \\
\hline
\ $I_{+}$ & $M\geq3 \cdot 10^{13}M_P$ \\
\hline
\ $I_{-}$ & $M\geq 10^{15}M_P$ \\
\hline
\ $II_{+}$ & $M\geq 500 M_P$ \\
\hline
\ $II_{-}$ & $M\geq 6 \cdot 10^{3}M_P$ \\
\hline
\end{tabular}
\end{center}

It is important to notice that these solutions confirm the intuitive 
expectations described in the introduction: in fact, the values of threshold
momenta are (in most cases) profoundly modified by the modifications 
of dispersion relations produced by violations of Lorentz Invariance.

\section { Theoretical Motivation }

The above analisys is quite general and does not refer to specific models.
It is however important to notice that there are, in the literature, 
models which lead violations of LI in the form discussed
above (see for instance Amelino-Camelia et al. 1997, Amelino-Camelia et al.
1998 and for a more complete list Aloisio et al. 2000).

More generally, {\it if} violations of LI are ascribed to Quantum 
Gravitational effects we can classify the violations in a general way. In fact
QG effects imply that the metric of space-time is non trivial when examined
near the Planck scale
\be
g_{\mu \nu} = \eta_{\mu \nu}+h_{\mu \nu}
\label{eq:l8}
\ee
where $\eta_{\mu \nu}=$ diag(1,-1,-1,-1) is the flat metrics and 
$h_{\mu \nu}$ is a term fluctuating in the vicinity of the Planck
length
\footnote{Notice that we cosider (8) as a phenomenological description.
In QG the concept of a background (flat) metric might be ill-posed.}.

Consider $g_{\mu \nu} P^{\mu}P^{\nu}$ in such a metric so that 
\be
 g_{\mu \nu} P^{\mu} P^{\nu} =
\eta_{\mu \nu} P^{\mu} P^{\nu} +
 h_{\mu \nu} P^{\mu} P^{\nu} 
\label{eq:l9}
\ee
A particle traveling with energy $E$ averages the fluctuations over
a scale $\lambda/l_P$ where $\lambda$ is its de-Broglie wavelength
($\lambda/l_P \approx 10^8$ for a $10^{20}$ eV proton) so that: 
\be
\left < g_{\mu \nu} P^{\mu} P^{\nu} \right >_{\lambda\over l_P}
=\eta_{\mu \nu} P^{\mu} P^{\nu} +
\left < h_{\mu \nu} P^{\mu} P^{\nu} \right >_{\lambda\over l_P} =m^2
\approx E^2-p^2 
+ ({l_p \over \lambda})^n \bar h_{\mu \nu} P^{\mu} P^{\nu}+...
\label{eq:l10} 
\ee
where the two parameters $n$, $\bar h_{\mu \nu}$ describe the 
(possible) violations of 
LI due to QG effects; clearly in the spirit of the discussion of Sect. 2,
if $n > 2 $ then the effects are in any case likely to be negligible
when $p<<M_P$.
On the other hand $\bar h$ may be:
\begin{enumerate}
\item
$\bar h_{\mu \nu} = 0$: 
the dispersion relation is not modified;
\item
$\bar h_{\mu \nu} \propto \eta_{\mu \nu} \Longrightarrow 
E^2-p^2 \propto f(m^2),  f(0)=0$: this gives the mildest violation,
that does not affect photon propagation. 
\item
$\bar h_{\mu \nu}$ non diagonal: 
$ E^2-p^2\neq m^2$. 
For instance:   $ \bar h_{ij}=\mp \delta_{ij},\bar h_{00}
=\bar h_{0i}=0$ \\  is a possible (among many others) choice that leads to 
$ E^2-p^2 = m^2 \pm p^2(p/M)^n+..$ and for $n=1 (2)$ generates 
the violations of type $I_\pm (II_\pm)$.
\end{enumerate}
Even in case 1, in general, one can have (Ford and Yu 1999)
$$\left < \bar h_{\mu \nu} \bar h_{\mu' \nu'} \right > \ne 0
\Longrightarrow \left < (E^2-p^2-m^2)^2 \right >_{\lambda\over l_P}
\ne 0$$ 
with possibly observable effects.

The vacuum of Quantum Gravity is often described as filled by 
virtual, Planck mass Black Holes (Hawking 1995).
Even if the full vacuum metric might be difficult to manage,
it is interesting to notice that 
in the field of a single BH the dipersion relation becomes: 
$E^2-p^2 = m^2 \pm {l_P \over L} (E^2+p^2)$ which holds when
$({l_P \over L}  << 1)$, and we expect $L \approx \lambda$. The average
over the ensemble of fluctuating BHs is non trivial, since it 
depends on the QG dynamics; it seems however natural to expect that
$\left < (E^2-p^2-m^2)^2 \right >^{1\over 2} 
\approx {\cal O}({p^3 \over M_P})$

\section { Conclusions}

Cosmic Ray experiments already in operation have the capability to 
investigate the validity of Lorentz Invariance to a high degree
of sensitivity. There are already a number of events above
the GZK cut-off and the situation will improve dramatically in a few 
years with the beginning of data taking of the Auger experiment.
And  $\gamma$-telescopes are already in the position of studying
the spectrum of a few (up to now) extargalactic sources up to tens of TeV.

These experiments  can test LI down 
to length scales in principle much smaller than the Planck length.
This is extremely important, since it means that we do not need Planck
energies to study the Planck physics if we appropriately chose the 
processes to study. 

The experimental situation is still rather unclear: no sign of the GZK 
cut-off has been seen up in the proton spectrum up to a few in $10^{20}$ eV
(Takeda et al 1999, Abu-Zayyad et al. 1999) 
while the 
$\gamma$ spectrum of Mkn501 although showing a bend, might be inconsistent
with expectations based on a new estimate of IR background
(Protheroe and Meyer 2000).

It is important however to use much caution in interpreting the experimental
data: our knowledge of the possible sources of highest energy CRs is 
rather poor, and in general it is certainly premature to 
invoke violations of LI to explain experimental data. 
This  might be true also in the case of 
VHE $\gamma$ astronomy, although, at least in the case of Mkn501,
also due to (almost) simultaneous
multi-wavelength observation (Guy et al. 2000), the  
experimental data are constraining the source spectrum. 
Again, the statistics is at 
present low, and there is uncertainty on the IR flux, but the situation
is likely to improve. And, $\gamma$-experiments performed in the
PeV range (Catanese etal., Ghia et al.), 
where the knowledge of the background is extremely better,
are possibly the best arena for testing Lorentz Invariance.

Due to the large cross-sections involved, 
this kind of experiments might not need to be only of cosmic nature.
In fact a terrestrial photon target, 
containing $10^{21}$ infrared ($\epsilon \approx 0.01$ eV) 
photons/cm$^3$ 
and 1 cm long, would have an 
efficiency of $10^{-4}$ to convert TeV photons into $e^+$ $e^-$ pairs.
This target does not seem unfeasible: a 1 W monochromatic source would
produce as many IR photons in one second, and  
if TeV photons could be produced (at LHC?) with sufficient intensity,
this device could test models of LI violation of type I up to the 
Planck scale. 

\section{Acknowledgements}
While the content of section 3 is responsibility of the present authors, 
this talk uses much of the material of Aloisio et al. (2000). We thank 
Piera L. Ghia and Pasquale Blasi for continuos collaboration. 
We thank also Angelo Galante for collaboration and Giuseppe Di
Carlo for discussions.

\section { References}

\reff
Abu-Zayyad T. et al., 1999 
Proc. 26th ICRC (Salt Lake City, USA), {\bf 3}, 264.

\reff
Aharonian F. A. et al., 1999, Astron. \& Astroph. {\bf 349}, 11.

\reff 
Aloisio, R., Blasi, P., Ghia, P.L. and Grillo, A.F.,
2000, Phys. Rev. {\bf D62}, 53010.

\reff
Amelino Camelia G., Ellis J.,  Mavromatos N.E. and Nanopoulos D.V., 1997,
Int. J. Mod. Phys. {\bf A12}, 2029.

\reff
Amelino Camelia G., Ellis J.,  Mavromatos N.E., Nanopoulos D.V. and Sarkar S.,
1998, Nature {\bf 393}, 763.

\reff
Amelino Camelia G. and Piran T., 2000, preprint hep-ph/0006210.

\reff
Catanese M. et al., 1996, Astrophys. J. {\bf 469}, 572.

\reff
Coleman S. and  Glashow S.L., 1997, Phys. Lett. {\bf B405}, 249.

\reff
Cronin J. W., 1992, Nucl. Phys. B (Proc. Suppl.) {\bf 28B}, 213.

\reff
Hawking S.W., 1995 Phys. Rev. {\bf D53}, 3099.

\reff
Ford L.H. and  Yu H., 1999, preprint gr-qc/9907037. 

\reff
Garay J.L., 1999, Int. J. Mod. Phys. {\bf A14}, 4079. 

\reff
Ghia P.L. et al., 1999, Nucl. Phys. B (Proc. Suppl.) {\bf 70}, 506.

\reff
Gonzalez-Mestres L., 1997, preprint physics/9702026.

\reff
Goldreich P. and Morrison P., 1964, Sov. Phys. - JETP {\bf 18}, 239.

\reff
Gould R.J. and Schreder G.P., 1966, Phys. Rev. Lett. {\bf 16}, 252.

\reff
Guy J., Renault C.,  Aharonian F.A., Rivoal M. and Tavernet J.P., 2000,
preprint astro-ph 0004355

\reff
Greisen K., 1966, Phys. Rev. Lett. {\bf 16}, 748.

\reff
Kirzhnits D.A. and Chechin V.A., 1971, Sov. Jour. Nucl. Phys. {\bf 15}, 585.

\reff
Nikishov A.I., 1962, Sov. Phys. - JETP {\bf 14}, 393.

\reff
Protheroe R.J. and Biermann P.L., 1996, Astropart. Phys. {\bf 6}, 45 ; 
1997, Erratum-ibid {\bf 7}, 181.

\reff
Protheroe R.J. and Meyer H., 2000, preprint astro-ph/0005349.

\reff
Stecker F.W., 1999, preprint astro-ph/9904416.

\reff
Takeda M. et al., 1999, Astrophys. J. {\bf 522}, 225.

\reff
Wheeler J.A., 1957, Ann. Phys. (N.Y.) {\bf 2}, 604.

\reff
Zatsepin G.T. and Kuzmin V.A., 1966, 
Pis'ma Zh. Ekps. Teor. Fiz. {\bf 4}, 114 [JETP Lett. {\bf 4}, (1966) 78].

\end{document}